\documentclass[iop]{emulateapj}  

\usepackage{color}
\usepackage{hyperref}  
\usepackage{breakurl}  

\newcommand\kms{\ifmmode{\rm km\thinspace s^{-1}}\else km\thinspace s$^{-1}$\fi}

\shortauthors{Torres}
\shorttitle{Gravity darkening in JKTEBOP}

\begin{document} 
\submitted{Accepted for publication in RNAAS}

\title{On the correct use of gravity darkening coefficients in the
  JKTEBOP eclipsing binary code}

\author{Guillermo Torres}

\affiliation{Center for Astrophysics $\vert$ Harvard \& Smithsonian,
  60 Garden St., Cambridge, MA 02138, USA; gtorres@cfa.harvard.edu}

\begin{abstract} 

Users of the {\tt JKTEBOP} code to solve the light curves of eclipsing
binaries often confuse the gravity darkening coefficients,
$y(\lambda)$, with the bolometric gravity darkening exponents,
$\beta$.  {\tt JKTEBOP} requires the wavelength-dependent
coefficients.  I show that the numerical values of $y(\lambda)$ and
$\beta$ can be rather different, leading to potential biases in the
solution if the wrong quantities are used.

\end{abstract}

\section{Introduction}
\label{sec:introduction}

\cite{vonZeipel:1924} showed that under conditions of radiative
equilibrium, the emergent bolometric flux at any point on the surface
of a tidally or rotationally distorted star is proportional to the
local gravity. A rotating star will be flattened at the poles, which
will then be brighter (hotter) than the equatorial regions. This ``gravity
darkening'' (or gravity brightening) effect can have a non-negligible
impact on the light curve of an eclipsing binary.

More generally, the gravity darkening law may be expressed as $T_{\rm
  eff}^4 \propto g^{\beta}$, where $g$ is the local gravity and
$T_{\rm eff}$ the effective temperature. For stars in radiative
equilibrium $\beta = 1.0$ \citep{vonZeipel:1924}, whereas
\cite{Lucy:1967} showed that stars with deep convective envelopes have
$\beta \approx 0.32$, on average. More recently \cite{Claret:1998}
published extensive tabulations giving $\beta$ as a function of mass,
age, and effective temperature.

Some light curve modeling programs including the Wilson-Devinney code
\citep{Wilson:1971, Wilson:1979}, {\tt WINK} \citep{Wood:1971,
  Wood:1973}, {\tt ELC} \citep{Orosz:2000}, and {\tt PHOEBE}
\citep{Prsa:2018} expect the user to supply a value for the exponent
$\beta$ of the bolometric gravity darkening law. Application of the
effect to the specific bandpass of the light curve is then dealt with
internally, using either theoretical model atmospheres or a blackbody
approximation.

Other light curve modeling programs adopt a different formulation to
deal with gravity darkening directly at the wavelength of the
photometric observations. Examples include {\tt ellc}
\citep{Maxted:2016}, as well as the {\tt EBOP} code \citep{Etzel:1981,
  Popper:1981} and its descendants, {\tt eb} \citep{Irwin:2011} and
      {\tt JKTEBOP} \citep{Southworth:2004, Southworth:2013}, which
      are most suitable for well-detached systems. These programs use
      a Taylor expansion of the wavelength-dependent flux as a
      function of local gravity, retaining only the linear term,
      consistent with other approximations in the binary model
      \citep[see, e.g.,][]{Kopal:1959, Kitamura:1983}. The coefficient
      of the linear term that these codes require is $y(\lambda)$,
      which is specific to the bandpass under consideration. Tables of
      $y(\lambda)$ based on model atmospheres, calculated for a wide
      range of stellar properties,
      have been published by \cite{Claret:2011} for several of
      the most common photometric systems, and more recently by
      \cite{Claret:2017} for the photometric band of NASA's
      planet-finding TESS mission \citep{Ricker:2014}.  The latter
      source of coefficients is particularly useful, as TESS has
      already collected high-precision light curves for many thousands
      of eclipsing binaries over much of the sky.

Among the simpler light curve programs using the linearized,
wavelength-specific formulation of gravity darkening, {\tt JKTEBOP} is
one of the more popular ones for its speed and ease of use. This Note
is motivated by the observation that some eclipsing binary studies
that use {\tt JKTEBOP} confuse the gravity darkening
\emph{coefficient} with the \emph{exponent} of the bolometric law, and
adopt numerical values for $\beta$ instead of $y(\lambda)$. The author
himself has made this mistake once or twice in the past. Other papers
using {\tt JKTEBOP} give no indication of what was assumed about
gravity darkening, so it is not possible to tell whether or not the
same mistake was made.

The use of the linearized prescription for gravity darkening is
clearly described in the original documentation for {\tt EBOP},
although admittedly that unpublished document is difficult to obtain.
As of this writing, full documentation for {\tt JKTEBOP} itself is not
yet unavailable, aside from some notes on the
website\footnote{\url{https://www.astro.keele.ac.uk/jkt/codes/jktebop.html}}
and in the sample input file, but the source code does correctly
indicate that the input for gravity darkening should be the
coefficient $y(\lambda)$. This was also explained in detail by
\cite{Torres:2017}, although mostly only in a footnote. It is also
described correctly in the documentation for the {\tt eb} code of
\cite{Irwin:2011}, which relies on the same underlying binary model as
     {\tt JKTEBOP}, although {\tt JKTEBOP} users would probably have
     little reason to consult that document, as it pertains to a
     different program.\footnote{The paper describing the {\tt ellc}
       code of \cite{Maxted:2016}, which also uses the linearized
       gravity darkening law, explicitly mentions the
       wavelength-dependent $y(\lambda)$ quantities, but confusingly
       calls them exponents.}  It is perhaps not surprising,
     therefore, that many authors are unaware of this detail, and
     proceed with their {\tt JKTEBOP} light curve solution adopting
     values for $\beta$ instead of $y(\lambda)$.

\section{Numerical differences between coefficients and exponents}
\label{sec:difference}

\begin{figure}[t!]
\epsscale{1.15}
\plotone{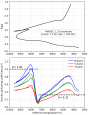}
\figcaption{\emph{Top:} Representative PARSEC isochrone by
  \cite{Chen:2014}.  \emph{Bottom:} Gravity darkening coefficients
  $y(\lambda)$ (for microturbulence $\xi = 2$~\kms) calculated along the
  isochrone in the top panel, for the Johnson $B$ and $V$ bands, and
  TESS $T$ \citep{Claret:2011, Claret:2017}.
  The dashed lines indicate the numerical values for the gravity darkening
  \emph{exponent} $\beta$ for stars with radiative and convective
  envelopes, which are sometimes adopted erroneously instead of the
  correct gravity darkening \emph{coefficients}
  $y(\lambda)$. \label{fig:fig1}}
\end{figure}

To illustrate the numerical differences between the bandpass specific
coefficients and the exponents of the gravity darkening law, I adopt
a representative solar-metallicity model isochrone from
\cite{Chen:2014}, shown in the top panel of Figure~\ref{fig:fig1}. For
each point along this isochrone, I extracted the $y(\lambda)$ values
from the tables of \cite{Claret:2011} and \cite{Claret:2017}, shown in
the bottom panel, which vary considerably with temperature and not in
a monotonic way.  I also indicate the numerical values of the
bolometric exponent often adopted for radiative stars \citep[$\beta =
  1.0$;][]{vonZeipel:1924} and convective stars \citep[$\beta =
  0.32$;][]{Lucy:1967}.

For stars of spectral type K ($\sim$4000--5000~K) the numerical values
of $y(\lambda)$ for the bluer wavelengths ($B$, $V$) are rather
different from the $\beta$ value typically adopted for such stars, by
up to a factor of two or more. Differences are smaller but still
significant in the TESS band, depending on the temperature. For hotter
stars, mistakenly adopting a value of $\beta = 1.0$ in using {\tt
  JKTEBOP}, instead of the proper wavelength-dependent coefficient
$y(\lambda)$, will be a poor approximation for all except the mid A
stars in the bluer bandpasses. Errors will be larger at longer
wavelengths.

It is difficult to predict how this mistake might bias the results of
a particular light curve solution with {\tt JKTEBOP}, or with other
programs that expect $y(\lambda)$ values instead of $\beta$. In
general the impact will depend on the binary configuration, the
stellar properties of the components, and the wavelength and
photometric precision of the observations. For well-detached systems
with nearly spherical stars (for which {\tt JKTEBOP} is most
suitable), the gravity darkening effect is small so using the wrong
number may not be serious. On the other hand, the TESS light curves
now available for large numbers of eclipsing binaries are much more
precise than ground-based photometry, so the impact may be more
noticeable. In any case, the correct gravity darkening values to use
with {\tt JKTEBOP} are the wavelength-dependent coefficients
$y(\lambda)$, and not the exponents $\beta$.

\begin{acknowledgements}

J.\ Southworth and A.\ Claret are thanked for a careful reading of the
manuscript.

\end{acknowledgements}

\end{document}